# Distributed Global Optimization (DGO)

Homayoun Valafar
Complex Carbohydrate
Research Center
University of Georgia
220 Riverbend Road
Athens, GA 30602
homayoun@mond1.ccrc.uga.edu

Okan K. Ersoy
School of Electrical
Engineering
Purdue University
West Lafayette, IN 47907
ersoy@ecn.purdue.edu

Faramarz Valafar
Complex Carbohydrate
Research Center
University of Georgia
220 Riverbend Road
Athens, GA 30602
faramarz@mond1.ccrc.uga.edu

*Abstract*

A new technique of global optimization and its applications in particular to neural networks are presented. The algorithm is also compared to other global optimization algorithms such as Gradient descent (GD), Monte Carlo (MC), Genetic Algorithm (GA) and other commercial packages.

This new optimization technique proved itself worthy of further study after observing its accuracy of convergence, speed of convergence and ease of use. Some of the advantages of this new optimization technique are listed below:

1. Optimizing function does not have to be continuous or differentiable.
2. No random mechanism is used, therefore this algorithm does not inherit the slow speed of random searches.
3. There are no fine-tuning parameters (such as the step rate of G.D. or temperature of S.A.) needed for this technique.
4. This algorithm can be implemented on parallel computers so that there is little increase in computation time (compared to linear increase) as the number of dimensions increases. The time complexity of $O(n)$ is achieved.

## *Introduction*

The new algorithm most resembles the Genetic Algorithm (GA)[1]. Like GA, Distributed Global Optimization (DGO) proceeds with the search of the optimal point by creating a population of points. Therefore DGO possesses most advantages of the GA. However unlike GA, DGO will not allow the unfit points to propagate to the next generation and thus the progression towards the optimal point is enhanced.

Another aspect of this algorithm resembles to that of the Monte Carlo (MC) method[5]. Like MC, DGO starts the search for the optimal point from several different starting points. By incorporating this characteristic, the search for the optimal point will have a certain element of randomness, which in term will cause a deeper optimal point. However since the number of random points is drastically smaller than MC method, DGO does not suffer from the slow convergence time as MC.

DGO starts the search for the optimal point from the parent point. This parent point is represented by n bits. From this parent 2n-1 new points will be generated by transforming the parent point in a deterministic fashion.

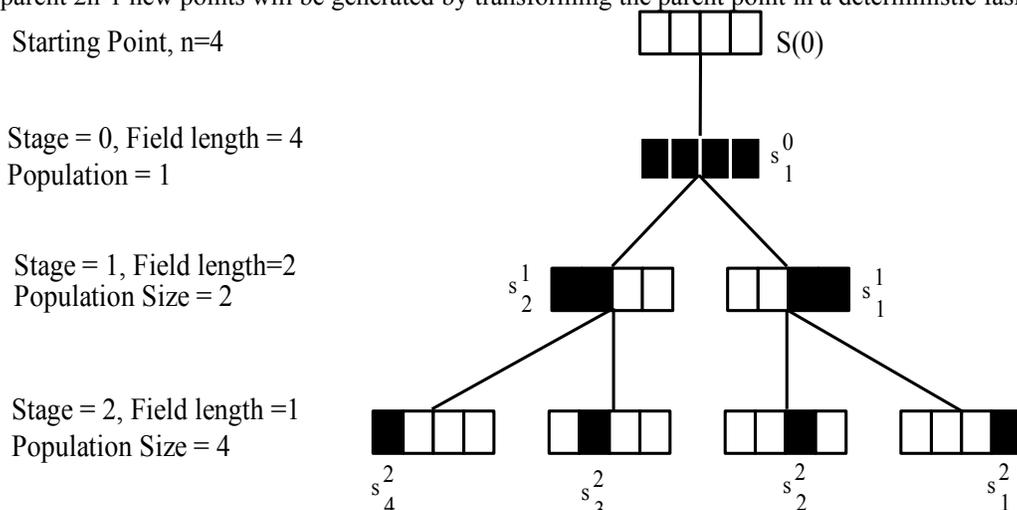

Figure 1 : An example of population generation for n=4.

DGO constructs and examines points in a tree like structure.  A graphical illustration of the creation of population through an entire cycle is shown in Figure 1 (In this example a vector length of 4 bits is assumed). In the above graph, $s_i^j$ indicates the i-th element of the j-th stage. Each $s_i^j$ is a possible candidate for the next state S(i+1).

## *Outline of the Algorithm*

To study the search of the algorithm in the feasible region of space, we need to establish a notation for the search. For this purpose, we denote the initial state of the algorithm as S(0). S(i) is referred to as the parent for the ith iteration. Furthermore, $s_i^j$ is the i-th element in the j-th stage

The general outline of the algorithm is as follows:
1. Select an initial parent string as either the user's choice or a random point. Evaluate the function at that point.
2. From the current parent generate 2n-1 children by transforming different segments of the parent string.
3. Among these points find the one with the lowest function value.
4. If there is a new minimum, then set the new string as S(i+1) (the new parent point) and go to 2.
5. If there is no deeper minimum, then increase the resolution.
6. If current resolution is less than maximum resolution, then go to 2, else terminate the algorithm.

Although in step 2 any qualified transformation can be used, in this experiment a gray code was used as the transformation . An effective transformation constructs the heart of this algorithm. For example, although binary inversion can be used as the transformation, the results of the optimization will not be effective. For more details on the gray code and inverse gray code transformations refer to [10]. In step 2 the transformation is constructed of the following steps:
1. Convert the entire binary string from two's complement to gray code.
2. Apply a binary inversion to the corresponding segment of the string.
3. Convert the string back to two's complement by the means of inverse-gray code transformation.

The above algorithm can be used for multidimensional problems as well by representing the set of all the independent variables as a single binary vector. Since all members of the population are created from one parent point, thus there are no dependencies in creation of each member. Conceptually, all members of a population can be created simultaneously. Because of this attribute of the algorithm, it can very easily be implemented on an SIMD machine for a total speedup of O(n).

## *Dynamic Resolution*

As it was shown above, from a parent string of length n, there will be 2n-1 children produced. From this new generation of points, one will be selected as the parent of the next generation. The number 2n-1 may be a large number of points to evaluate and therefore time consuming. One alternative to this approach is to start the search for the optimal point with a smaller resolution and gradually increase the resolution. The specifics of the algorithm are listed below:
1. Search the space for the optimal point using n bit resolution, for example n=8.
2. Once the optimal point is found, append n random bits to the right end of the result for a total of 2n bits.
3. Using the new starting point, continue the search for convergence.
4. Continue the above 3 steps in further runs to reach a desired resolution.

This modification has two obvious advantages. The first advantage is the total convergence time. The second effect of dynamic resolution is the ability to escape local optimal point. The random attachment of the bits at the lower half of each word, allows a certain degree of randomness throughout the search for the optimal point. This randomness will help the algorithm to escape from any local optimal point.

## *Multiple Starting Points*

One method of breaking away from a casual trappings in a local minimum is to start the search from several random starting points. This randomness  is integrated into the new algorithm by simply starting the search from several random initial points. By doing so, the strengths of the Monte Carlo method is integrated into this algorithm.

Although, in general, majority of algorithms are capable of the above integration, in practice this integration is not desirable because of the additional execution time. In practice, the new algorithm proved itself fast enough to afford multiple runs in order to find a deeper minimum.

## *Comparison to Other Algorithms*

This algorithm inherits most characteristics of the two resembling algorithms. However this inheritance is only with respect to the better qualities of the genetic algorithm and the Monte Carlo method. Each one of these methods have certain short comings which were eliminated in the new algorithm. Some of the shortcomings of GA and the MC method which are eliminated are as follows:

### Genetic Algorithm
1. Lack of a convergence criterion. Although the GA is capable of finding the global minimum, there is no simple method of ceasing the search.
2. Slow progression toward the optimal point due to random properties of the algorithm. This problem is common among all non-deterministic algorithms.
3. Since some of the less desirable points may propagate from one generation to the next, a slower and less effective convergence towards the optimal point results.
4. For an optimal performance, the size of the population need to be predetermined. By selecting a small number, not enough gene variation may be present, whereas by selecting a large population size, each stage of evolution will be very time consuming and slow.

### Monte Carlo Method
1. For this algorithm to find the global optimal point with a probability of P, means that P*100 percent of the points in the feasible region need to be examined. This number grows exponentially as the number of dimensions grows and therefore is very time consuming.
2. Like GA, this algorithm does not have any convergence criterion.

In contrast, DGO has the following characteristics:
1. DGO incorporates a distinct convergence and stopping rule.
2. Since the winner among the current generation will propagate to the next generation, there is no possibility for propagation of defective elements to the next generations.
3. Because of the deterministic property of the transformation, there are no random elements present in this algorithm.
4. Unlike GA there is no need for a predetermined population size. Population size is determined based on the problem size.
5. Since the new algorithm is fast enough in a single run for finding a final point, it can afford to be run multiple number of times.

## *Example of Movement Through Space*

The Gray code transformation plays a very important role in the performance of this algorithm. To test the Gray code transformation further, a more difficult function with more local minima was tested. The results are shown in figure 2 below. There are several number of minima present in this function, and therefore is a good test for falling into a local optimal point.

$$f_1(x)=\sin(x)+\sin(10x/3)+\ln(x)-0.84x \quad 2.7 \leq X \leq 7.5$$

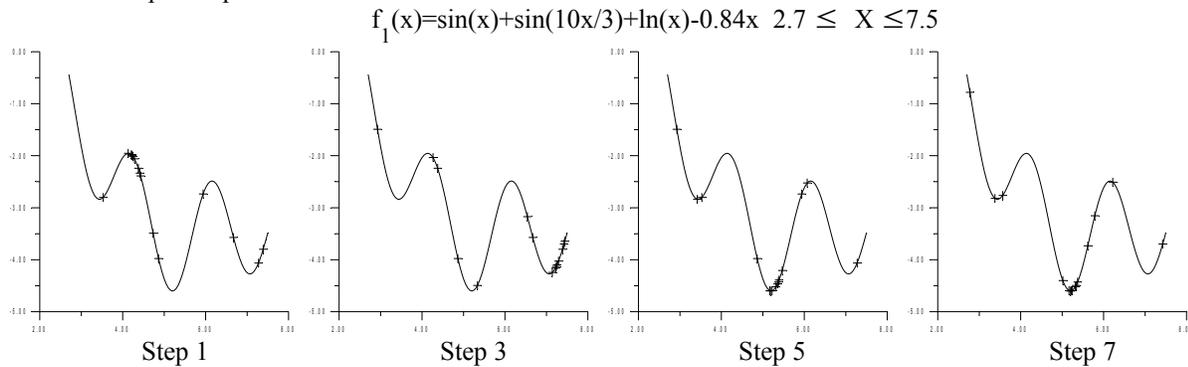

Step 1    Step 3    Step 5    Step 7

Figure 2. An example of the progression through the space using the Gray code transformation.

## *One and Two Dimensional Functions*

Test functions for the one and two dimensional spaces are given in [1,2,7]. These test functions represent practical problems well. For comparison purposes the results of the new algorithm were compared to that of fmin (a function of matlab), Eureka , gradient descent[2,11], genetic algorithm and simulated annealing[3,4]. Since the last three techniques required a very specific and delicate parameters, or their performance severely depended on the starting point, their results were not successful in one and two dimensional spaces. Figures 3, 4, 5 and 6 display examples of functions tested. The final results are shown in Tables 1 and 2. For more information on test functions please refer to [10].

It may be worth mentioning that in the worst case, the execution time of DGO did not exceed 1 second with 32 bit representation on an IBM 486DLC/40 compatible, running at 40 MHz. The original source code was written in C without any code optimization.

|  | Optimal point | Eureka | Fmin | DGO |
|---|---|---|---|---|
| $f_1(x)$ | x=5.1997785 | x=5.1997787 | x=5.1998 | x=5.1997785 |
| $f_2(x)$ | x=17.039199 | x=5.3622486 | x=5.3622 | x=17.039198 |
| $f_3(x)$ | $x^3$=5.7917947 | x=1.1062827 | x=7.3892 | x=5.7917947 |
| $f_4(x)$ | x=-0.67958096 | x=7.5639089 | x=0.6796 | x=-0.67958096 |
| $f_5(x)$ | x=0.0 | x=-4.9406E-9 | x=-4.4934 | x=0.00000001 |

Table 1 Results of Different Algorithms for One Dimensional Problems.

|  | True Minimum | Eureka | Fmins | DGO |
|---|---|---|---|---|
| $f_1(x,y)$ | (0.08983,-0.7126) | x=0.089842286<br>y=-0.7126564 | x=0.0899<br>y=-0.7126 | x=0.08983<br>y=-0.7126 |
| $f_2(x,y)$ | (0,0) | x=7.8464E-8<br>y=9.438E-8 | [-10 10] -> [-10.9041 -10.9041]<br>[-1 1] -> [-0.7772 0.7772]<br>[-5 5] -> [-4.4934 4.4935] | x=+/-0.0000076<br>y=+/-0.0000076 |
| $f_3(x,y)$<br>Figure 5 | (0,0) | x=1.00000<br>y=1.00000 | [-1 1] -> [-0.9910 0.99113]<br>[-0.9 0.9] -> [-0.9911 0.9911]<br>[-0.7 0.7] -> [-0.9911 0.9911]<br>[-0.6 0.6] -> [-0.9911 0.9911] | x=0.000015<br>y=0.000015 |
| $f_4(x,y)$<br>Figure 6 | (0,0) | x=0.99108311<br>y=0.99108313 | [-0.9 0.9] -> [-1.0407 1.0407]<br>[0.8 0.8] -> [0.6939 0.6939]<br>[0.5 0.5] -> [0 0.3469]<br>[0.3 0.3] -> [0.3469 0.3469] | x=0.000015<br>y=0.000015 |

Table 2 Results of Different Algorithms for Two Dimensional Problems.

$f_2(x)=\sin(x) + \sin(2x/3)$  $3.1 \leq X \leq 20.4$   $f_3(x)=-\sin(2x+1)-\sin(3x+2)-\sin(4x+3)-\sin(5x+4)-\sin(6x+5)$ $-10 \leq X \leq 10$

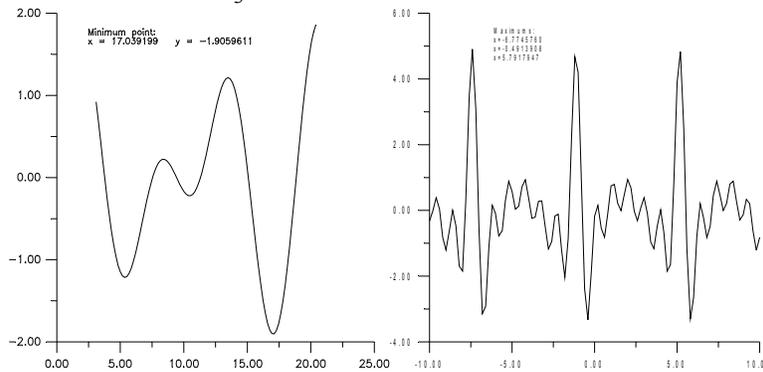

Figure 3.                                    Figure 4.

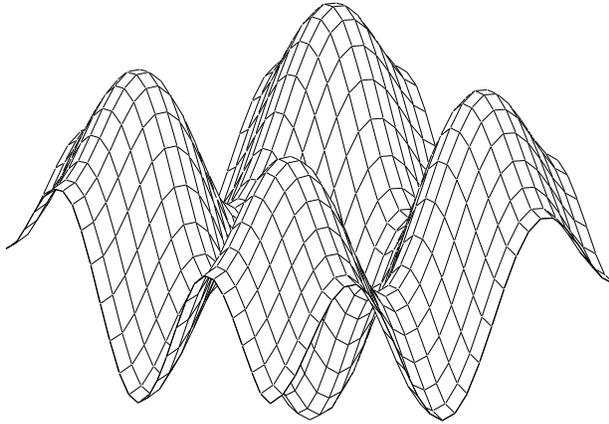 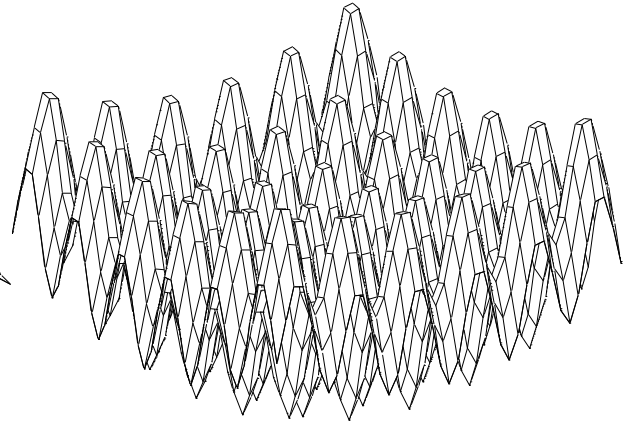

Figure 5. Figure 6.

## Neural Network Applications

Although formulated test functions give some insight to the operation of a certain algorithm, the true test of success remains in the real world applications. Artificial Neural Networks[8,9,10,11] (ANN) is one example of real life application which demands an effective optimization application. As a real test of DGO, it was applied to an ANN solving XOR problem as well as 8 class remote sensing application.

XOR problem contains 2 inputs and 1 output. The network used was 2 input, 2 hidden and 1 output neuron. Although this problem appears simple, it has many local optimal points. The existence of the numerous optimal points makes it a relatively difficult optimizing algorithm. For more information on this problem refer to [11]. Table 3 and figures 7,8 demonstrate the results of this problem.

The data set for the 8 class remote sensing experiment is based on Multispectral Earth observational remote sensing data called Flight Line C1 (FLC1). This multispectal image was collected with an airborne scanner in June 1966 at noon time. The FLC1 consists of 12 band signals. Each band signal represents a gray level image of that spectral band. Each point of one spectral image represents one of 256 gray levels. From the 12 band signals only the 8 dominant classes of the data were used. This problem required 688 variables. For more information on this problem refer to[9,12].

In figures 9 and 10 the results of Back Propagation and DGO are compared. As it can be seen the BP is trapped in a minimum and is not able to break away. However the DGO manages to fall into a much deeper minimum and continue the search. The starting point for DGO was 6000 and after 4 hours of execution time (clock time), the error had decreased to 1400 in comparison to 48 hours of execution for BP.

|  | BackPropagation | Monte Carlo | DGO |
|---|---|---|---|
| Square Error | Error = 0.02 | Error = 0.0139 | Error = 0.0000029 |

Table 3. Table of Results for Exclusive-or Problem.

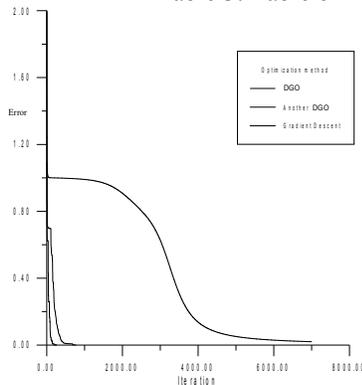 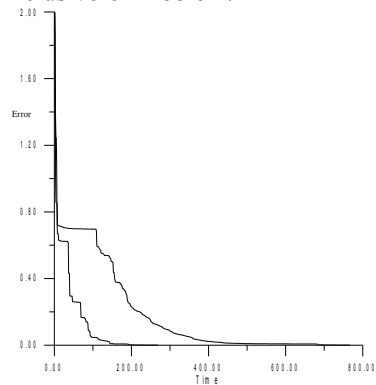

Figure 7. Error trace of DGO and Gradient Descent.   Figure 8. Error trace for DGO

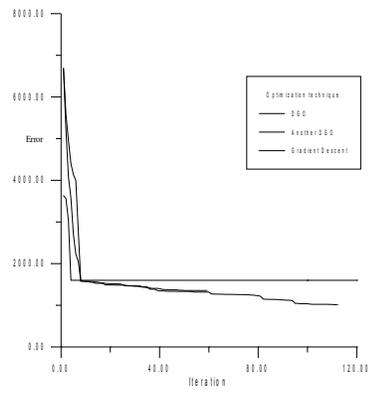 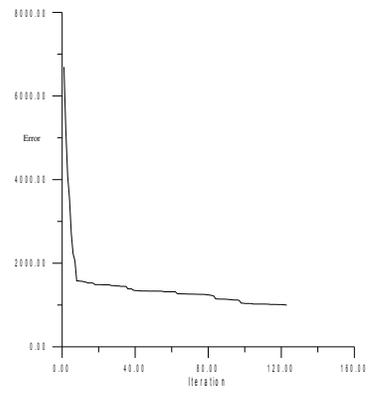

Figure 9. Error trace of DGO versus Gradient Descent.   Figure 10. Error trace for DGO alone.

## *CONCLUSIONS*
While some algorithms reach an execution time of exponential complexity, DGO does no worse than $O(n^2)$. Another advantage of the DGO is that it has a space complexity of $O(n)$. The DGO has been found to perform satisfactorily with artificial as well as practical functions. This algorithm has managed to successfully combine speed and accuracy into one package. One of the other attributes of DGO is its capacity for parallelization. As the problem size grows (because of the higher resolution or larger number of dimensions) more number of function evaluations are needed. This could be considered very costly. This effect can be countered by distributing the function evaluations across the PEs of a parallel processor. Since parallel machines with a number of processors are becoming more widely available, it would be justified to aim our goals towards the increase of the number of function evaluations at the cost of distributing the evaluations across PEs to increase performance without the increase in computation time. At this time the performance of DGO has been tested in problems with as many as 110,000 variables. Finally, the summary of the characteristics of DGO are listed below:
1. No fine tuning parameters which will alter the performance of the algorithm are required.
2. The algorithm works well with 1, 2 and 4-dimensional spaces as well as large dimensional spaces.
3. The algorithm is easily parallelizable.
4. Implementation of the algorithm (programming) is easy.
5. The algorithm has been observed to be more immune from falling into local minima and it often results in the global minimum.
6. Since there are no random mechanisms used in this algorithm, the search seems to be very fast.